\documentclass[prodmode,acmtist]{acmsmall} 
\pdfoutput=1

\usepackage[ruled]{algorithm2e}
\usepackage{listings}
\usepackage{xcolor}
\usepackage{graphicx}
\usepackage{multirow}
\usepackage{url}

\definecolor{purple}{rgb}{0.375,0,1}

\newcommand{\hide}[1]{}
\newcommand{\specialcell}[2][c]{%
  \begin{tabular}[#1]{@{}l@{}}#2\end{tabular}}
\newcommand{\xhdr}[1]{\vspace{2mm}\noindent{{\bf #1.}}}

\acmVolume{8}
\acmNumber{1}
\acmArticle{00}
\acmYear{2016}
\acmMonth{0}

\doi{0000000.0000000}

\issn{1234-56789}

\AtBeginDocument{\setlength{\tempdimen}{\textwidth}}
\begin{document}

\markboth{J. Leskovec and R. Sosi\v{c}}{SNAP: A General Purpose Network Analysis and Graph Mining Library}

\title{SNAP: A General Purpose Network Analysis and Graph Mining Library}
\author{JURE LESKOVEC
\affil{Stanford University}
ROK SOSI\v{C}
\affil{Stanford University}}



\begin{abstract}
Large networks are becoming a widely used abstraction
	for studying complex systems in a broad set of disciplines,
	ranging from social network analysis
	to molecular biology and neuroscience.
Despite an increasing need to analyze and manipulate large networks,
	only a limited number of tools are available for this task.

Here,
	we describe Stanford Network Analysis Platform (SNAP),
	a general-purpose, high-performance system
	that provides easy to use, high-level operations
		for analysis and manipulation of large networks.
We present SNAP functionality, describe its implementational details,
	and give performance benchmarks.
SNAP has been developed for single big-memory machines and
it balances the trade-off between
	maximum performance,
	compact in-memory graph representation, and
	the ability to handle dynamic graphs
		where nodes and edges are being added or removed over time.
SNAP can process massive networks with hundreds of millions of nodes
	and billions of edges.
SNAP offers over 140 different graph algorithms that can efficiently manipulate large graphs,
	calculate structural properties,
	generate regular and random graphs,
	and handle attributes and meta-data on nodes and edges.
Besides being able to handle large graphs,
	an additional strength of SNAP is
	that networks and their attributes are fully dynamic,
	they can be modified during the computation at low cost.
SNAP is provided as an open source library in C++ as well as a module in Python.

We also describe the Stanford Large Network Dataset, a set of social and information real-world networks and datasets, which we make publicly available.
The collection is a complementary resource to our SNAP software and is widely used for development and benchmarking of graph analytics algorithms.
\end{abstract}



\acmformat{Jure Leskovec and Rok Sosi\v{c}, 2016. SNAP: A General Purpose Network Analysis and Graph Mining Library.}

\begin{bottomstuff}
This work has been supported in part by
DARPA XDATA,
DARPA SIMPLEX,
NIH U54EB020405,
IIS-1016909,
CNS-1010921,
IIS-1149837, Boeing, and Stanford Data Science Initiative.

Author's addresses: J. Leskovec {and} R. Sosi\v{c}, Computer Science Department, 353 Serra Mall, Stanford University, Stanford, CA 94305.
\end{bottomstuff}

\maketitle

\maketitle


\section{Introduction}

The ability to analyze large networks is fundamental to study of complex systems in many scientific disciplines~\cite{easley2010networks,jackson2008social,newman2010networks}. With networks, we are able to capture relationships between entities, which allows us to gain deeper insights into the systems being analyzed~\cite{newman2003structure}. This increased importance of networks has sparked a growing interest in network analysis tools~\cite{batagelj1998pajek,hagberg2008exploring,kyrola2012graphchi,malewicz2010pregel}.

Network analysis tools are expected to fulfill a set of requirements. They need to provide rich functionality, implementing a wide range of graph and network analysis algorithms. Implementations of graph algorithms must be able to process graphs with 100s of millions of nodes. Graphs need to be represented in a compact form with a small memory footprint, since many algorithms are bound by the memory throughput. Powerful operators are required for modifying graph structure, so that nodes and edges in a graph can be added or removed, or new graphs can be constructed from existing ones. Additionally for a wide system adoption, it is desirable that the source code is available under an open source license.

While there has been significant amount of work on systems for processing and analyzing large graphs, none of the existing systems fulfills the requirements outlined above.
In particular, research on graph processing in large-scale distributed environments~\cite{gonzalez2012powergraph,malewicz2010pregel,kang2009pegasus,salihoglu2013gps,xin2013graphx} provides efficient frameworks, but these frameworks only implement a handful of most common graph algorithms, which in practice is not enough to make these tools useful for practitioners. Similarly, there are several user-friendly libraries that implement dozens of network analysis algorithms~\cite{batagelj1998pajek,csardi2006igraph,gregor2005parallel,hagberg2008exploring,o2005analysis}.
However, the limitations of these systems are that they might not scale to large graphs, can be slow, hard to use, or do not include support for dynamic networks. Thus, there is a need for a system that addresses those limitations and provides reasonable scalability, is easy to use, implements numerous graph algorithms, and supports dynamic networks.

Here, we present Stanford Network Analysis Platform (SNAP), which was specifically built with the above requirements in mind. SNAP is a general-purpose, high-performance system that provides easy to use, high-level operations for analysis and manipulation of large networks. SNAP has been developed for single big-memory multiple-cores machines and as such it balances the trade-off between maximum performance, compact in-memory graph representation, and the ability to handle dynamic graphs where nodes and edges are being added or removed over time.

SNAP offers methods that can efficiently manipulate large graphs, calculate structural properties, generate regular and random graphs, and handle attributes on nodes and edges. 
Besides being able to handle large graphs, an additional strength of SNAP is that network structure and attributes are fully dynamic, they can be modified during the computation via low cost operations.

Overall, SNAP implements 8 graph and network types, 20 graph generation methods/models, 20 graph manipulation methods, and over 100 graph algorithms, which provides in total over 200 different functions. It has been used in a wide range of applications, such as network inference~\cite{rodriguez2014uncovering}, network optimization~\cite{hallac2015network}, information diffusion~\cite{jure09memes,suen2013nifty}, community detection~\cite{jaewon14ieee}, and geo-spatial network analysis~\cite{leskovec2014geospatial}.
SNAP is provided for major operating systems as an open source library in C++ as well as a module in Python.
It is released under the BSD open source license and can be downloaded from \texttt{http://snap.stanford.edu/snap}.

Complementary to the SNAP software, we also maintain public Stanford Large Network Dataset Collection, an extensive set of social and information networks with about 80 different network datasets. The collection includes online social networks with rich dynamics and node attributes, communication networks, scientific citation networks, collaboration networks, web graphs, Internet networks, online reviews, as well as social media data. The network datasets can be obtained at \texttt{http://snap.stanford.edu/data}.

The remainder of the paper is organized as follows.
We discuss related graph analysis systems in Section~\ref{sec:related}.
The next two sections describe key principles behind SNAP.
We give an overview of basic graph and network classes in SNAP in Section~\ref{sec:basic}, while Section~\ref{sec:methods} focuses on graph methods.
Implementational details are discussed in Section~\ref{sec:implementation}.
An evaluation of SNAP and comparable systems with benchmarks on a range of graphs and graph algorithms is presented in Section~\ref{sec:benchmarks}.
Next, in Section~\ref{sec:data}, we describe Stanford Large Network Dataset Collection and, in Section~\ref{sec:documentation}, SNAP documentation and its distribution license.
Section~\ref{sec:conclusion} concludes the paper.

\section{Related Network Analysis Systems}
\label{sec:related}

In this section we briefly survey related work on systems for processing, manipulating, and analyzing networks. We organize the section into two parts. First, we discuss single-machine systems and then proceed to discuss how SNAP relates to distributed systems for graph processing.

One of the first single-machine systems for network analysis is
	Pajek~\cite{batagelj1998pajek},
	which is able to analyze networks with up to ten million nodes.
Pajek is written in Pascal and
	is distributed as a self-contained system with its own GUI-based interface.
It is only available as a monolithic Windows executable, and
	thus limited to the Windows operating system.
It is hard to extend Pajek with additional functionality
	or use it as a library in another program.
Originally, networks in Pajek
        are represented using doubly linked lists~\cite{batagelj1998pajek} and
while linked lists make it easy to insert and delete elements,
        they can be slow to traverse on modern CPUs,
        where sequential access to memory is much faster than random access.

Other widely used open source network analysis libraries
	that are similar in functionality to SNAP are
	NetworkX~\cite{hagberg2008exploring} and
	iGraph~\cite{csardi2006igraph}.
NetworkX is written in Python and
	implements a large number of network analysis methods.
In terms of the speed vs. flexibility trade-off,
        NetworkX offers maximum flexibility at the expense of performance.
Nodes, edges and attributes in NetworkX are represented by hash tables,
        called dictionaries in Python.
Using hash tables for all graph elements allows for maximum flexibility,
        but imposes performance overhead in terms of
                a slower speed and a larger memory footprint than
                alternative representations.
Additionally, since Python programs are interpreted,
        most operations in NetworkX take significantly longer time and
        require more memory than alternatives in compiled languages.
Overall, we find SNAP to be one to two orders of magnitude faster than NetworkX, while also using around 50 times less memory.
This means that, using the same hardware, SNAP can process networks that are 50 times larger or networks of the same size 100 times faster.

Similar to NetworkX in functionality but very different in implementation is the iGraph package~\cite{csardi2006igraph}.
iGraph is written in the C programming language and can be used as a library.
In addition, iGraph also provides interfaces for Python and R programming languages.
In contrast to NetworkX,
        iGraph emphasizes performance at the expense of the flexibility of the underling graph data structure.
Nodes and edges are represented by vectors and
        indexed for fast access and iterations over nodes and edges.
Thus graph algorithms in iGraph can be very fast.
However, iGraph's representation of graphs is heavily optimized for fast execution of algorithms that operate on a static network. As such, iGraph is prohibitively slow when making incremental changes to the graph structure, such as node and edge additions or deletions.
Overall, we find SNAP uses about three times less memory than iGraph due to extensive use of indexes in iGraph, while being about three times slower executing a few algorithms that benefit from indexes and fast vector access. 
However, the big difference is in flexibility of the underlying graph data structure. For example, SNAP was five orders of magnitude faster than iGraph in our benchmarks of removal of individual nodes from a graph.

While SNAP was designed to work on a single large-memory machine, an alternative approach would be to use a distributed system to perform network analysis.
Examples of such systems include Pregel~\cite{malewicz2010pregel}, PowerGraph~\cite{gonzalez2012powergraph}, Pegasus~\cite{kang2009pegasus}, and GraphX~\cite{xin2013graphx}.
Distributed graph processing systems can in principle process larger networks
	than a single machine,
	but are significantly harder to program, and
	more expensive to maintain.
Moreover, none of the existing distributed systems comes with a large suite of graph processing functions and algorithms. Most often, graph algorithms, such as community detection or link prediction, have to be implemented from scratch.

We also note a recent trend where, due to decreasing RAM prices,
	the need for distributed graph processing systems has diminished in the last few years.
Machines with large RAM of 1TB or more have become relatively inexpensive.
Most real-world graphs comfortably fit in such machines,
	so multiple machines are not required to process them~\cite{perez15ringo}.
Multi-machine environments also impose considerable execution overhead
	in terms of communication and coordination costs,
	which further reduces the benefit of distributed systems.
A single machine
	thus provides an attractive platform for graph analytics~\cite{perez15ringo}.

\section{SNAP Foundations}
\label{sec:basic}

SNAP is a system for analyzing graphs and networks.
In this section we shall provide an overview of SNAP, starting by introducing some basic concepts.
In SNAP we define {\it graphs} to consist of a set of nodes and a set of edges,
	each edge connecting two nodes.
Edges can be either {\it directed} or {\it undirected}.
In {\it multigraphs}, more than one edge can exist between a pair of nodes.
In SNAP terminology {\it networks} are defined as graphs, where attributes or features, like ``age'', ``color'', ``location'', ``time'' can be associated with nodes as well as edges of the network.

SNAP is designed in such a way that graph/network methods are agnostic to the underling graph/network type/representation. As such most methods work on any type of a graph/network. So, for most of the paper we will be using terms {\it graphs} and {\it networks} interchangeably,	meaning {\it graph and/or network} and the specific meaning will be evident from the context.

An alternative terminology to the one we use here is to use the term {\it graph} to denote mathematical objects and the term {\it network} for real-world instances of graphs, such as an online social network, a road network, or a network of protein interactions. However, inside the SNAP library we use the terminology where graphs represent the ``wiring diagrams'', and networks are graphs with data associated with nodes and edges.

\subsection{Graph and Network Containers}
\label{sec:containers}

SNAP is centered around core foundational classes that store graphs and networks.
We call these classes {\it graph and network containers}.
The containers provide several types of graphs and networks,
        including directed and undirected graphs, multigraphs, and
        networks with node and edge attributes.
In order to optimize for execution speed and memory usage,
	an application can chose the most appropriate container class
        so that critical operations are executed efficiently.

An important aspect of containers is that
	they all have a unified interface for accessing the graph/network structure as well as for traversing nodes and edges.
This common interface is used by {\it graph methods}
	to implement more advanced graph algorithms.
Since the interface is the same for all graph and network containers,
	these advanced methods in SNAP are generic in a sense that each method can work on a container of any type.
Implementation of new algorithms is thus simplified as
	each method needs to be implemented only once and can then be executed on any type of a graph or a network.
At the same time, the use of SNAP library is also streamlined.
It is easy to substitute one type of graph container for another
	at the container creation time,
	and the rest of the code usually does not need to be changed.

\begin{figure}[t]
\centering
\includegraphics[width=0.75\columnwidth]{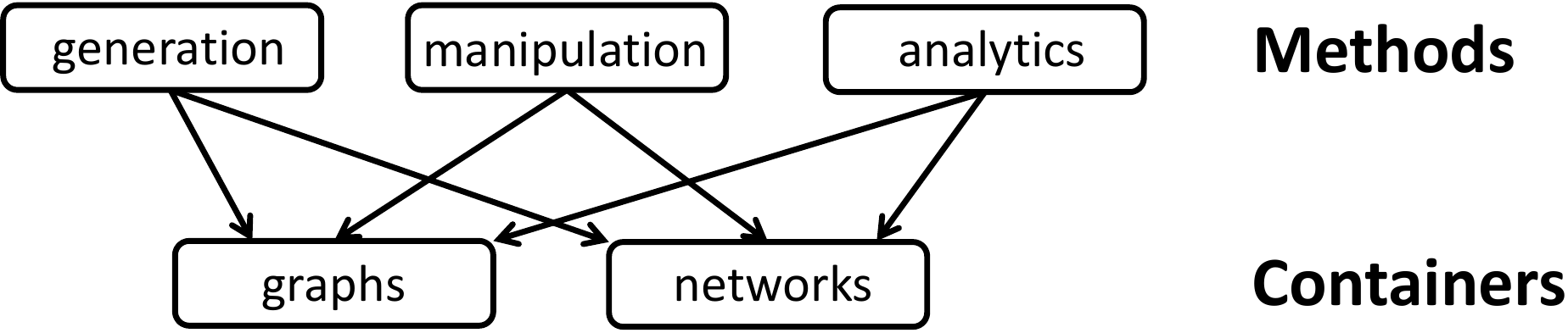}
\caption{SNAP components: graph and network containers and methods.}
\label{figure:components}
\end{figure}

Methods that operate on graph/network containers can be split into several groups (Figure~\ref{figure:components}):
graph {\it generation} methods which create new graphs as well as networks,
graph {\it manipulation} methods which manipulate the graph structure, and
graph {\it analytic} methods which do not change the underlying graph structure, but compute specific graph statistics.
Graph methods are discussed further in Section~\ref{sec:methods}.

\begin{table}[t]
\centering
\tbl{SNAP Graph and Network Containers.
}{
        \setlength{\tabcolsep}{4pt}
        \begin{tabular}{l l}
        \hline
	{\bf Graph Containers} & \\
        \hline
	\rule{0pt}{2ex} 
	\texttt{TUNGraph} & Undirected graphs \\
	\rule{0pt}{2ex} 
	\texttt{TNGraph} & Directed graphs \\
	\rule{0pt}{2ex} 
	\texttt{TNEGraph} & Directed multigraphs \\
	\rule{0pt}{2ex} 
	\texttt{TBPGraph} & Bipartite graphs \\
        \hline 
	{\bf Network Containers} & \\
        \hline
	\rule{0pt}{2ex} 
	\texttt{TNodeNet} & \specialcell[t]{Directed graphs with node attributes} \\
	\rule{0pt}{2ex} 
	\texttt{TNodeEDatNet} & \specialcell[t]{Directed graphs with node and edge attributes} \\
	\rule{0pt}{2ex} 
	\texttt{TNodeEdgeNet} & \specialcell[t]{Directed multigraphs with node and edge attributes} \\
	\rule{0pt}{2ex} 
	\texttt{TNEANet} & \specialcell[t]{Directed multigraphs with dynamic node and edge attributes} \\
        \hline
        \end{tabular}
}
        \label{table:snap_classes}
\end{table}

Table~\ref{table:snap_classes} describes the multiple graph and network containers provided by SNAP. Each container is optimized for a particular type of graph or network.

Graph containers are \texttt{TUNGraph}, \texttt{TNGraph}, \texttt{TNEGraph},
	and \texttt{TBPGraph},
	which correspond to
	undirected graphs where edges are bidirectional,
	directed graphs where edges have direction,
	directed multigraphs where multiple edges can exist
		between a pair of nodes, and
	bipartite graphs, respectively.
Network containers are \texttt{TNodeNet},
	\texttt{TNodeEDatNet},
	\texttt{TNodeEdgeNet}, and
	\texttt{TNEANet},
	which correspond to 
	directed graphs with node attributes,
	directed graphs with node and edge attributes,
	directed multigraphs with node and edge attributes and
	directed multigraphs with dynamic node and edge attributes, respectively.

In all graph and network containers, nodes have unique identifiers (ids), which are non-negative integers. Node ids do not have to be sequentially ordered from one to the number of nodes,
but can be arbitrary non-negative integers. The only requirement is that each node has a unique id. In simple graphs edges have no identifiers and can be accessed by providing an pair of node ids that the edge connects. However, in multigraphs each edge has a unique non-negative integer id and edges can be accessed either by providing an edge id or a pair of node ids.

The design decision to allow arbitrary node (and edge) ids is important as it allows us to preserve node identifiers as the graph structure is being manipulated. For example, when extracting a subgraph of a given graph, the node as well as edge ids get preserved.

Network containers, except \texttt{TNEANet},
	require that types of node and edge attributes
	are specified at compile time.
These attribute types are simply passed as template parameters in C++,
	which provides a very efficient and convenient way to implement
	networks with rich data on nodes and edges.
Types of node and edge attributes in the \texttt{TNEANet} container
	can be provided dynamically,
	so new node and edge attributes can be added or removed at run time.

Graph and network containers vary in
	how they represent graphs and networks internally,
	so time and space trade-offs can be optimized
	for specific operations and algorithms.
Further details on representations are provided in Section~\ref{sec:implementation}.

\subsection{Functionality of Graph Containers}
\label{sec:iterators}

Container interface allows that the same commonly used primitives
	are used by containers of all types.
This approach results in significant reduction
	of the effort needed to provide new graph algorithms in SNAP,
	since most algorithms need to be implemented only once and
	can then be used for all the graph and network container types.

Common container primitives are shown in Table~\ref{table:snap_common}.
\begin{table}[t]
\centering
\setlength{\tabcolsep}{4pt}
\tbl{Common Graph and Network Methods.}
{
        \begin{tabular}{l l}
        \hline
	{\bf Nodes} &  \\ \hline
	\rule{0pt}{2ex} \texttt{AddNode} & Adds a node \\
	\rule{0pt}{2ex} \texttt{DelNode} & Deletes a node \\
	\rule{0pt}{2ex} \texttt{IsNode} & Tests, if a node exists \\
	\rule{0pt}{2ex} \texttt{GetNodes} & Returns the number of nodes \\
        \hline
	{\bf Edges} &  \\ \hline
	\rule{0pt}{2ex} \texttt{AddEdge} & Adds an edge \\
	\rule{0pt}{2ex} \texttt{DelEdge} & Deletes an edge \\
	\rule{0pt}{2ex} \texttt{IsEdge} & Tests, if an edge exists \\
	\rule{0pt}{2ex} \texttt{GetEdges} & Returns the number of edges \\
        \hline
	{\bf Graph Methods} &  \\ \hline
	\rule{0pt}{2ex} \texttt{Clr} & Removes all nodes and edges \\
	\rule{0pt}{2ex} \texttt{Empty} & Tests, if the graph is empty \\
	\rule{0pt}{2ex} \texttt{Dump} & Prints the graph in a human readable form \\
	\rule{0pt}{2ex} \texttt{Save} & Saves a graph in a binary format to disk \\
	\rule{0pt}{2ex} \texttt{Load} & Loads a graph in a binary format from disk \\
        \hline
	{\bf Node and Edge Iterators} &  \\ \hline
	\rule{0pt}{2ex} \texttt{BegNI} & Returns the start of a node iterator \\
	\rule{0pt}{2ex} \texttt{EndNI} & Returns the end of a node iterator \\
	\rule{0pt}{2ex} \texttt{GetNI} & Returns a node (iterator) \\
	\rule{0pt}{2ex} \texttt{NI++}  & Moves the iterator to the next node \\
	\rule{0pt}{2ex} \texttt{BegEI} & Returns the start of an edge iterator \\
	\rule{0pt}{2ex} \texttt{EndEI} & Returns the end of an edge iterator \\
	\rule{0pt}{2ex} \texttt{GetEI} & Returns an edge (iterator) \\
	\rule{0pt}{2ex} \texttt{EI++}  & Moves the iterator to the next edge \\
        \hline
        \end{tabular}
}
        \label{table:snap_common}
\end{table}
These provide basic operations for graph manipulation.
For example, they include primitives that add or delete nodes and edges,
	and primitives that save or load the graph.

Expressive power of SNAP comes from iterators
	that allow for a container independent traversal of nodes and edges.
Listing~\ref{code:snap_iterators} illustrates the use of iterators
	by providing examples of how all the nodes and edges in the graph can be traversed.

\begin{lstlisting}[caption={Iterating over Nodes and Edges. Top example prints out the ids and out-degrees of all the nodes. Bottom example prints out all the edges as pairs of edge source node id and edge destination node id. These traversals can be executed on any type of a graph/network container.},
  label=code:snap_iterators,xleftmargin=.06\textwidth,
  float=t]
// traverse all the nodes, print out-degree for each node
for (TNGraph::TNodeI NI=Graph->BegNI(); NI<Graph->EndNI(); NI++) {
  printf("node %d, outdegree %d\n", NI.GetId(), NI.GetOutDeg());
}

// traverse all the edges, print source and destination nodes
for (TNGraph::TEdgeI EI=Graph->BegEI(); EI<Graph->EndEI(); EI++) {
  printf("edge (%d, %d)\n", EI.GetSrcNId(), EI.GetDstNId());
}
\end{lstlisting}

The iterators are used consistently and extensively throughout the SNAP code base.
As a result,
	existing graph algorithms in SNAP
	do not require any changes in order to be applied
		to new graph and network container types.

Special attention has been paid in SNAP to performance of graph load and save operations.
Since large graphs with billions of edges can take a long time to load or save,
	it is important that these operations are as efficient as possible.
To support fast graph saving and loading operations,
	SNAP can save graphs directly in a binary format,
	which avoids a computationally expensive step
		of data serializing and deserializing.


\section{Graph Methods}
\label{sec:methods}

\begin{table*}[t!]
\small
\begin{center}
\tbl{Graph generators in SNAP.
}{
\begin{tabular}{ll}
\hline
{\bf Category} & {\bf Graph Generators} \\
\hline
{\it Regular graphs} & Complete graphs, circles, grids, stars, and trees; \\
{\it Basic random graphs} & \specialcell[t]{
	Erd\H{o}s-R\'{e}nyi graphs,
	Bipartite graphs, \\
	Graphs where each node has a constant degree, \\
	Graphs with exact degree sequence;
	} \\
{\it Advanced graph models} & \specialcell[t]{
	Configuration model~\cite{bollobas1980probabilistic}, \\
	Ravasz-Barabasi model~\cite{ravasz2003hierarchical}, \\
	Copying model~\cite{kumar2000stochastic}, \\
	Forest Fire model~\cite{leskovec2005graphs}, \\
	Geometric preferential model~\cite{flaxman2006geometric}, \\
	Barabasi-Albert model~\cite{barabasi1999emergence}, \\
	Rewiring model~\cite{milo2003uniform}, \\
	R-MAT~\cite{chakrabarti2004r}, \\
	Graphs with power-law degree distribution, \\
	Watts-Strogatz model~\cite{watts1998collective}, \\
	Kronecker graphs~\cite{leskovec2010kronecker}, \\
	Multiplicative Attribute Graphs~\cite{kim2012multiplicative}.
	} \\
\hline
\end{tabular}
}
  \label{table:snap_gen}
\end{center}
\end{table*}

\begin{table*}[t!]
\small
\begin{center}
\tbl{Graph manipulation and analytics methods in SNAP.
}{
\begin{tabular}{ll}
\hline
{\bf Category} & {\bf Graph Manipulation and Analytics} \\
\hline
{\it Graph manipulation} &
	\specialcell[t]{Graph rewiring, decomposition to connected\\ components, subgraph extraction, graph type\\ conversions;} \\
{\it Connected components} & 
	\specialcell[t]{Analyze weakly, strongly, bi- and 1-connected\\ components;} \\
{\it Node connectivity} &
	\specialcell[t]{Node degrees, degree distribution, in-degree,\\ out-degree, combined degree, Hop plot, Scree plot;} \\
{\it Node centrality algorithms} &
	\specialcell[t]{PageRank, Hits, degree-, betweenness-, closeness-,\\ farness-, and eigen-centrality, personalized PageRank;} \\
{\it Triadic closure algorithms} &
	\specialcell[t]{Node clustering coefficient, triangle counting, clique\\ detection;} \\
{\it Graph traversal} &
	\specialcell[t]{Breadth first search, depth first search, shortest\\ paths, graph diameter;} \\
{\it Community detection} &
	\specialcell[t]{Fast modularity, clique percolation, link clustering,\\Community-Affiliation Graph Model, BigClam, CoDA,\\ CESNA, Circles;} \\
{\it Spectral graph properties} &
	\specialcell[t]{Eigenvectors and eigenvalues of the adjacency matrix,\\spectral clustering;} \\
{\it K-core analysis} &
	\specialcell[t]{Identification and decomposition of a given graph to\\ $k$-cores;} \\
{\it Graph motif detection} &
	Counting of small subgraphs; \\
{\it Information diffusion} &
	Infopath, Netinf; \\
{\it Network link and node prediction} &
	Predicting missing nodes, edges and attributes. \\
\hline
\end{tabular}
}
  \label{table:snap_methods}
\end{center}
\end{table*}

SNAP provides efficient implementations of
	commonly used traditional algorithms for graph and network analysis,
	as well as recent algorithms that employ machine learning techniques on graph problems,
	such as community detection~\cite{yang2013overlapping,jaewon14ieee,julian14circles},
	statistical modeling of networks~\cite{kim2012multiplicative,kim2013nonparametric},
	network link and missing node prediction~\cite{kim2011network},
	random walks~\cite{lofgren2016ppr},
	network structure inference~\cite{gomez2010inferring,gomez2013structure}.
These algorithms have been developed within our research group or
	in collaboration with other groups.
They use SNAP primitives extensively and
	their code is made available as part of SNAP distributions.

Graph methods can be split into the following groups:
	graph creation, graph manipulation, and graph analytics.
Graph creation methods, called generators, are shown in Table~\ref{table:snap_gen}.
They implement a wide range of models
	for generation of regular and random graphs,
	as well as graphs that model complex real-world networks.
Table~\ref{table:snap_methods} shows
	major families of graph manipulation and analytics methods.
Next, we describe advanced graph methods in more detail.

\subsection{Community Detection}

Novel SNAP methods for community detection are based on the observation
	that overlaps between communities in the graph are more densely connected
	than the non-overlapping parts of the communities~\cite{jaewon14ieee}.
This observation matches empirical observations in many real-world networks,
	however, it has been ignored by most traditional community detection methods.

The base method for community detection is the Community-Affiliation Graph Model (AGM)~\cite{yang2012community}.
This method has been extended in several directions to cover
	networks with millions of nodes and edges~\cite{yang2013overlapping},
	networks with node attributes~\cite{YanMcALes13}, and 
	2-mode communities~\cite{YanMcALes14}.

Community-Affiliation Graph Model
	identifies communities in the entire network.
SNAP also provides a complementary approach to network wide community detection.
The Circles method~\cite{McALes12}
	uses the friendship network connections
		as well as user profile information
	to categorize friends from a person's ego network
		into social circles~\cite{julian14circles}.

\subsection{Predicting Missing Links, Nodes, and Attributes in Networks}

The information we have about a network might often be partial and incomplete,
	where some nodes, edges or attributes
		are missing from the available data.
Only a subset of nodes or edges in the network is known,
	the rest of the network elements are unknown.
In such cases, we want to predict the unknown, missing network elements.

SNAP methods for these prediction tasks are based on
	the multiplicative attribute graph (MAG) model~\cite{kim2012multiplicative}.
The MAG model can be used to predict
	missing nodes and edges~\cite{kim2011modeling},
	missing node features~\cite{kim2012latent}, or
	network evolution over time~\cite{kim2013nonparametric}.

\subsection{Fast Random Walk Algorithms}

Random walks can be used to determine the importance or authority of nodes in a graph. In personalized PageRank, we want to identify important nodes from the point of view of a given node~\cite{benczur2005spamrank,lofgren2014fast,PageRank}.

SNAP provides a fast implementation of the problem of computing personalized PageRank scores for a distribution of source nodes to a given target node~\cite{lofgren2016ppr}. In the context of social networks, this problem can be interpreted as finding a source node that is interested in the target node. The fast personalized PageRank algorithm is birectional. First, it works backwards from the target node to find a set of intermediate nodes 'near' it and then generates random walks forwards from source nodes in order to detect this set of intermediate nodes and compute a provably accurate approximation of the personalized PageRank score.

\subsection{Information Diffusion}

Information diffusion and virus propagation are fundamental network processes.
Nodes adopt pieces of information or become infected and
	then transmit the information or infection to some of their neighbors.
A fundamental problem of diffusion over networks is the problem of {\em network inference}~\cite{gomez2010inferring}.
The network inference task is to use node infection times in order to reconstruct the transmissions as well as the network that underlies them.
For example, in an epidemic,
	we can usually observe just a small subset of nodes being infected, and
	we want to infer the underlying network structure over which the epidemic spread.

SNAP implements an efficient algorithm for network inference, where the problem is to find the optimal network that best explains a set of observed information propagation cascades~\cite{gomez12netinf}.
The algorithm scales to large datasets and
	in practice gives provably near-optimal performance.
For the case of dynamic networks,
	where edges are added or removed over time and
		we want to infer these dynamic network changes,
	SNAP provides an alternative algorithm~\cite{gomez2013structure}.


\section{SNAP Implementation Details}
\label{sec:implementation}

SNAP is written in the C++ programming language and optimized for compact graph representation while preserving maximum performance.
In the following subsections we shall discuss implementational details of SNAP.

\subsection{Representation of Graphs and Networks}
\label{sec:graph_representation}

Our key requirement when designing SNAP was that
	data structures are flexible in allowing for
		efficient manipulation of the underlying graph structure,
	which means that adding or deleting nodes and edges
		must be reasonably fast and not prohibitively expensive.
This requirement is needed, for example, for the processing of dynamic graphs,
	where graph structure is not known in advance, and
	nodes and edges get added and deleted over time.
A related use scenario is motivated by on-line graph algorithms,
	where an algorithm incrementally modifies existing graphs
	as new input becomes available.

Furthermore, we also want our algorithms to offer high performance and
	be as fast as possible given the flexibility requirement.
These opposing needs of flexibility and high performance
	pose a trade-off between
	graph representations that allow for efficient structure manipulation and
	graph representations that are optimized for speed.
In general,
	flexibility is achieved by using hash table based representations,
	while speed is achieved by using vector based representations.
An example of the former is NetworkX~\cite{hagberg2008exploring},
	an example of the latter is iGraph~\cite{csardi2006igraph}.

\begin{figure}[t]
        \centering
\begin{minipage}[b]{0.25\textwidth}
	\includegraphics[width=0.90\textwidth]{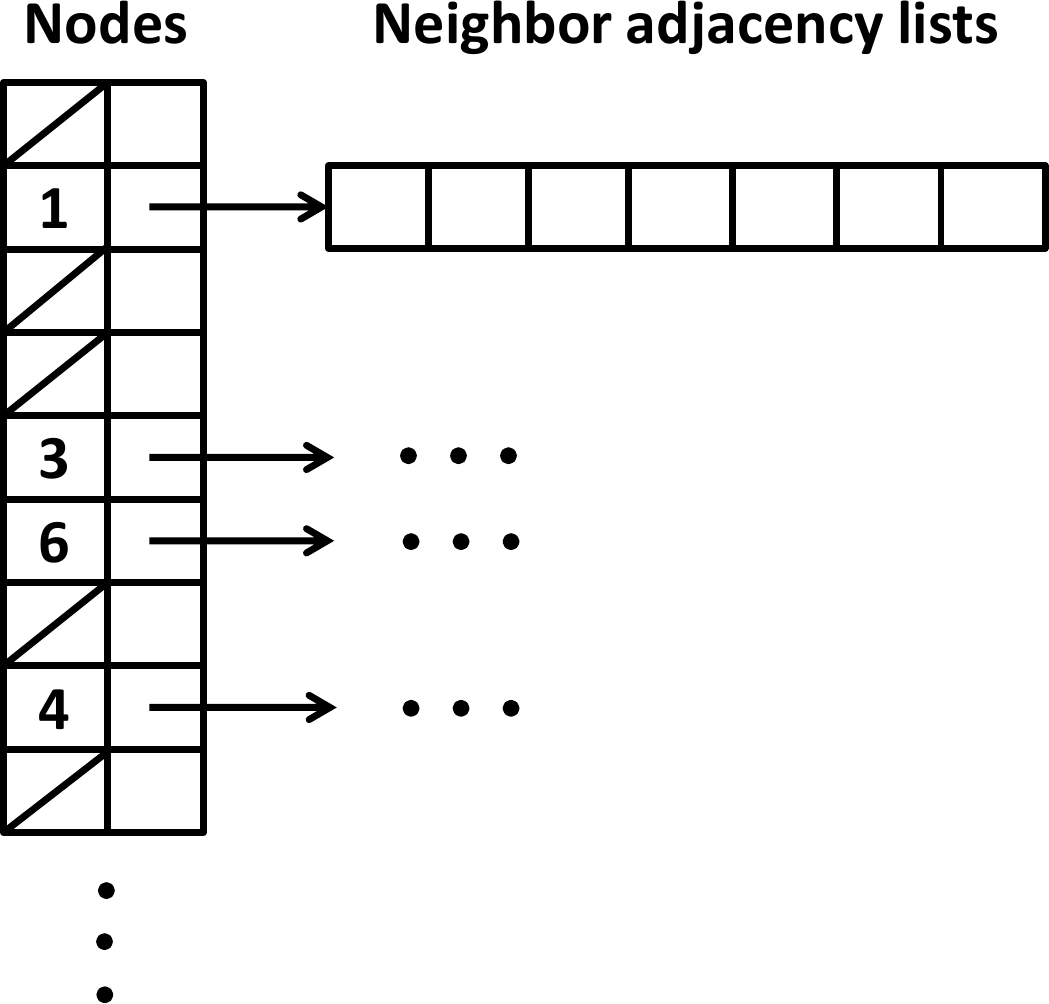}
	\subcaption{Undirected graph}
	\label{fig:undirected}
\end{minipage}
\begin{minipage}[b]{0.30\textwidth}
	\includegraphics[width=0.90\textwidth]{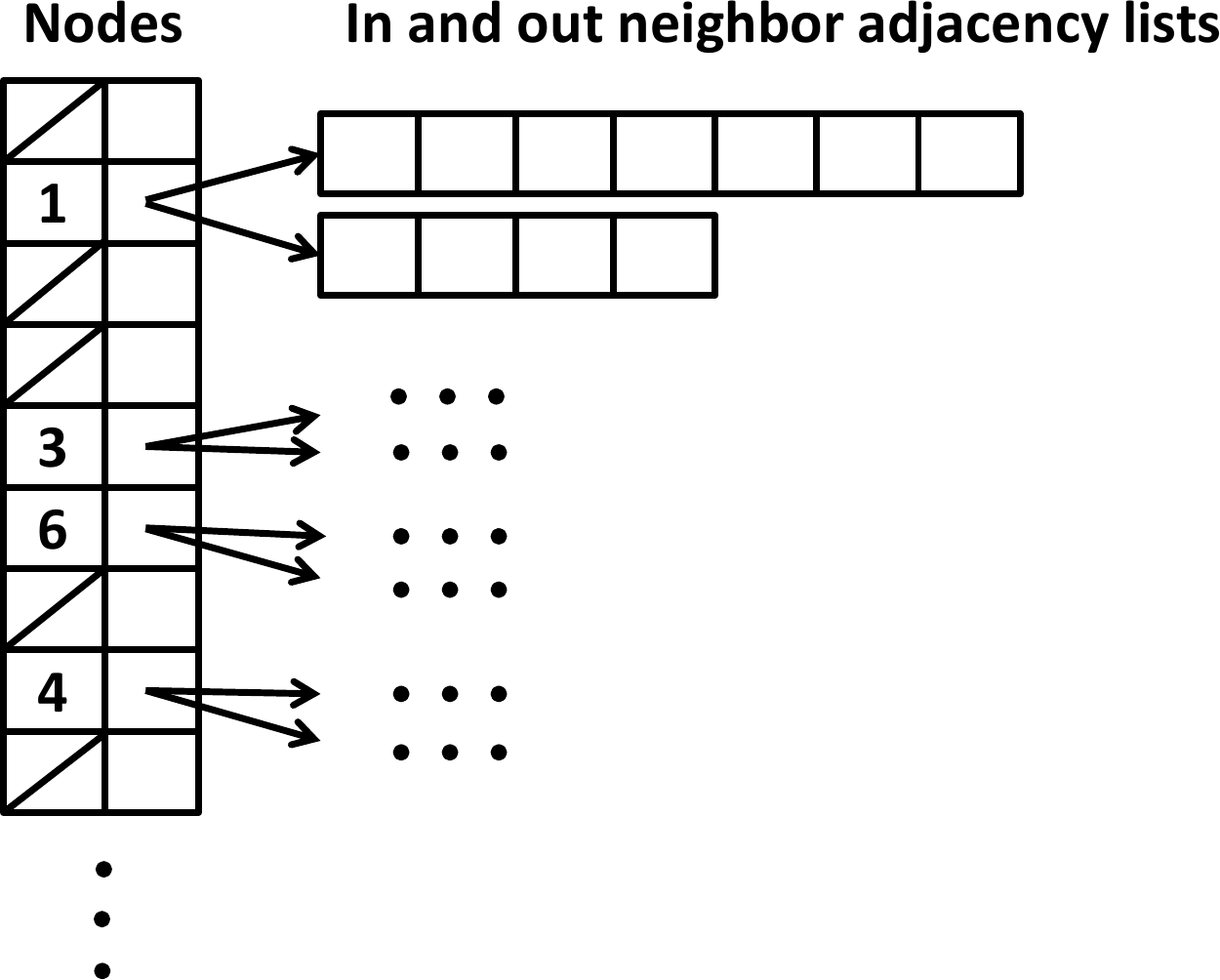}
	\subcaption{Directed graph}
	\label{fig:directed}
\end{minipage}
\begin{minipage}[b]{0.375\textwidth}
	\includegraphics[width=0.90\textwidth]{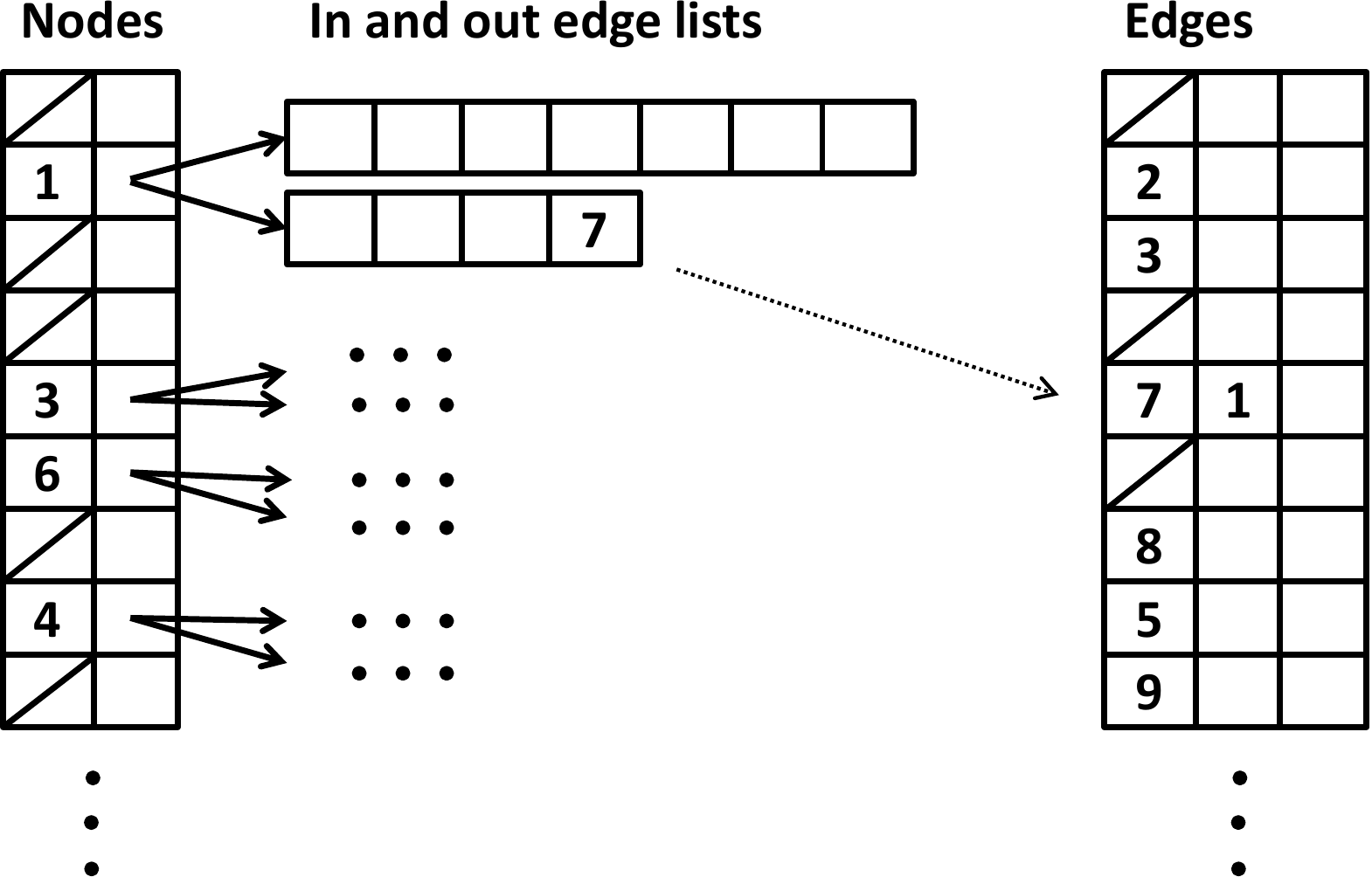}
	\subcaption{Directed multi-graph}
	\label{fig:multi}
\end{minipage}
        \caption{A diagram of graph data structures in SNAP. Node ids are stored in a hash table, and each node has one or two associated vectors of neighboring node or edge ids.}
	\label{fig:representation}
\end{figure}

\xhdr{SNAP graph and network representation}
For SNAP, we have chosen a middle ground
	between all-hash table and all-vector graph representations.
A graph in SNAP is represented
	by a hash table of nodes in the graph.
Each node consists of a unique identifier and
	one or two vectors of adjacent nodes,
		listing nodes that are connected to it.
Only one vector is used in undirected graphs,
	while two vectors,
		one for outgoing and another one for incoming nodes/edges,
	are used in directed graphs.
In simple graphs,
	there are no explicit edge identifiers,
	edges are treated as pairs of a source and a destination node instead.
In multigraphs,
	edges have explicit identifiers,
		so that two edges between the same pair of nodes
		can be distinguished.
An additional hash table is required in this case for the edges,
	mapping edge ids to the source and destination nodes.
Figure~\ref{fig:representation} summarizes graph representations in SNAP.

The values in adjacency vectors are sorted for faster access.
Since most of the real-world networks are sparse
	with node degrees significantly smaller than
		the number of nodes in the network,
	while at the same time exhibiting
		a power law distribution of node degrees,
	the benefits of maintaining the vectors in a sorted order
		significantly outweigh the overhead of sorting.
Sorted vectors also allow for fast and ordered traversal and
	selection of node's neighbors,
	which are common operations in graph algorithms.

As we show in experiments (Section~\ref{sec:benchmarks}),
	SNAP graph representation also optimizes memory usage for large graphs.
Although it uses more memory for storing nodes than some alternative representations,
	it requires less memory for storing edges.
Since a vast majority of relevant networks have more edges than nodes,
	the overall memory usage in SNAP is smaller than representations
		that use less memory per node but more per edge.
A compact graph representation is important for handling very large networks,
	since it determines the sizes of networks that can be analyzed
		on a computer with a given amount of RAM.
With a more compact graph representation and smaller RAM requirements,
	larger networks can fit in the RAM available and can thus be analyzed.
Since many graph algorithms are bound by memory throughput,
	an additional benefit of using less RAM to represent graphs is that
	the algorithms execute faster,
	since less memory needs to be accessed.

\xhdr{Time complexity of key graph operations}
Table~\ref{table:snap_complexity} summarizes time complexity of
	key graph operations in SNAP.
\begin{table}
\small
\begin{center}
\tbl{Time complexity of key graph operations in SNAP. $deg_{max}$ denotes the maximum node degree in the graph.}
{
\begin{tabular}{ll}
\hline
{\bf Operation} & {\bf Time Complexity} \\
\hline
Get node, get next node & $O(1)$ \\
Get edge, get next edge & $O(1)$ \\
Add, delete, test an existence of a node & $O(1)$ \\
Add, delete an edge & $O(deg_{max})$ \\
Test an existence an edge & $O(log(deg_{max}))$ \\
\hline
\end{tabular}
}
  \label{table:snap_complexity}
\end{center}
\end{table}
It can be seen that most of the operations complete in constant time of $O(1)$,
	and that the most time consuming are edge operations,
	which depend on the node degree.
However, since most of the nodes in real-life networks have low degree,
	edge operations overall still perform faster than alternative approaches.
One such alternative approach is to maintain neighbors in a hash table
	rather than in a sorted vector.
This alternative approach does not work well in practice,
	because hash tables are faster than vectors
		only when the number of elements stored is large.
But most nodes in real-time networks have a very small degree,
	and hash tables will be slower than vectors for these nodes.
We find that a small number of large degree nodes does not compensate
	for the time lost with a large number of small degree nodes.
Additionally,
	an adjacency hash table would need to be maintained for each node,
	leading to significantly increased complexity
		with hundreds of millions of hash tables for graphs
			with hundreds of millions of nodes.

As we show in the experimental section (Section~\ref{sec:benchmarks}),
	the representation of graphs in SNAP is able to provide
		high performance and compact memory footprint,
	while allowing for efficient additions or deletions of nodes and edges.

\subsection{Implementation Layers}

\begin{figure}
\centering
\includegraphics[width=0.70\columnwidth]{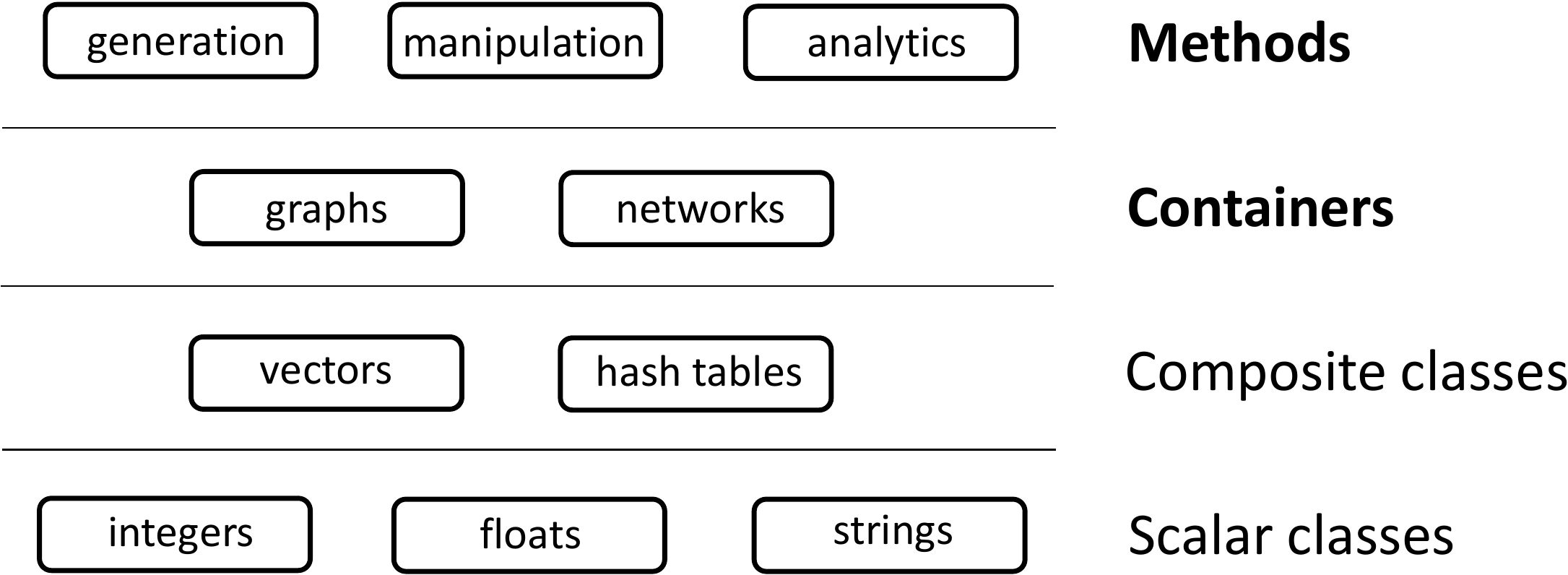}
\caption{Different layers of SNAP design.}
\label{figure:layers}
\end{figure}

SNAP is designed to operate in conceptual layers (see Figure~\ref{figure:layers}).
Layers are designed in such a way that every level abstracts out the complexity of the lower level.
The bottom layer comprises of basic scalar classes, like integers, floats, and strings.
Next layer implements composite data structures, like vectors and hash tables.
A layer above them are graph and network containers.
And the last layer contains graph generation, manipulation, and analytics methods.
SNAP implementation takes advantage of GLib,
	a general purpose C++ STL-like library (Standard Template Library),
	developed at Jo\v{z}ef Stefan Institute in Ljubljana, Slovenia.
GLib is being actively developed and used in
	numerous academic and industrial projects.

\xhdr{Scalar classes}
This foundational layer implements basic classes,
	such as integers, floating point numbers, and strings.
A notable aspect of this layer is
	its ability to efficiently load and save object instances
		to a secondary storage device.
SNAP saves objects in a binary format,
	which allows loading and storing of objects
		without any complex parsing and thus
	can be done at close to disk speeds.

\xhdr{Composite classes}
The next layer implements composite classes on top of scalar classes.
Two key composite classes are
	vectors, where elements are accessed by an integer index, and
	hash tables, where elements are accessed via a key.
The elements and keys in hash tables can have an arbitrary type.
SNAP expands fast load and save operations
	from scalar classes to vectors and hashes,
	so that these composite classes can be manipulated efficiently as well.

\xhdr{Graph and network containers}
The layer above vectors and hash tables are graph and network containers.
These were discussed in detail
	in Sections~\ref{sec:containers} and \ref{sec:graph_representation}.

\xhdr{Graph and network methods}
The top layer of SNAP implements graph and network algorithms.
These rely heavily on node and edge iterators,
	which provide a unified interface to all graph and network classes in SNAP (Section~\ref{sec:iterators}).
By using iterators, 
	only one implementation of each algorithm is needed
		to provide the algorithm for all the graph/network containers.
Without a unified iterator interface,
	a separate algorithm implementation would be needed for each container type,
	which would result in significantly larger development effort and
		increased maintenance costs.

For example, to implement a $k$-core decomposition algorithm~\cite{batagelj02cores}, 
one would in principle need to keep 
a separate implementation for each graph/network type (i.e., graph/network container).
However, in SNAP all graph/network containers expose the same set of functions and interfaces to access the graph/network structure. In case of the $k$-core algorithm, we need functionality to traverse all of the nodes of the network (we use node iterators to do that), determine the degree of a current node, and then delete it. All graph/network containers in SNAP expose such functions and thus a single implementation of the $k$-core algorithm is able to operate on any kind of graph/network container (directed and undirected graphs, multigraphs as well as networks).

\xhdr{Memory management}
In large software systems, memory management is an important aspect.
All complex SNAP objects, from composite to network classes,
	employ reference counting,
	so memory for an object is automatically released,
		when no references are left that point to the object.
Thus, memory management is completely transparent to the SNAP user and
	has minimal impact on performance,
	since the cost of reclaiming unused memory
		is spread in small chunks over many operations.


\section{Benchmarks}
\label{sec:benchmarks}

In this section, we compare SNAP with existing network analytics systems.
In particular, we contrast the performance of SNAP with two systems that are most similar in functionality, NetworkX~\cite{hagberg2008exploring} and iGraph~\cite{csardi2006igraph}. 

NetworkX and iGraph are single machine, single thread graph analytics libraries that occupy two opposite points in the performance vs. flexibility spectrum.
iGraph is optimized for performance,
	but not flexible in a sense that it supports 
    primarily only static graph structure (dynamically adding/deleting nodes/edges is prohibitively expensive).
	On the other hand, NetworkX is optimized for flexibility
		at the expense of lower performance.
SNAP lies in-between,
	providing flexibility while maximizing performance.

Furthermore, we also give a summary of our experiments with parallel versions of several SNAP algorithms~\cite{perez15ringo}. These experiments demonstrate that a single large-memory multi-core machine provides an attractive platform for the analysis of all-but-the-largest graphs. In particular, we show that performance of SNAP on a single machine measures favorably when compared to distributed graph processing frameworks.

All the benchmarks were performed on a computer with
	2.40GHz Intel Xeon E7-4870 processors and
	sufficient memory to hold the graphs in RAM.
Since all the systems are non-parallel,
	benchmarks utilized only one core of the system.
All benchmarks were repeated 5 times and the average times are shown.

\subsection{Memory Consumption}

A memory requirement to represent graphs is an important measure
	of a graph analytics library.
Many graph operations are limited by available memory access bandwidth,
	and a smaller memory footprint allows for faster algorithm execution.

To determine memory consumption,
	we use undirected Erd\H{o}s-R\'{e}nyi random graphs,
		$G(n,m)$, where $n$ represents the number of nodes, and
		$m$ the number of edges in the graph.
We measure memory requirements for $G(n,m)$ graphs
         at three different sizes
                $G(1M,10M)$, $G(1M,100M)$, and $G(10M,100M)$, where $1M$ denotes $10^6$.
We have chosen those graph sizes
	to illustrate system scaling
		as the number of nodes or the average node degree increases.

\begin{table}[h]
\centering
\tbl{Memory requirements of undirected Erd\H{o}s-R\'{e}nyi random graphs, the $G(n,m)$ model. Memory usages are in MB. Overall, SNAP uses three times less memory than iGraph and over 40 times less memory than NetworkX.}
{
\begin{tabular}{r r | r r r}
        \hline
        \multicolumn{2}{c|}{Graph size} & \multicolumn{3}{c}{Memory usage [MB]}  \\
        Nodes & Edges & SNAP & iGraph & NetworkX  \\
        \hline
	$1M$ & $10M$   &   137 &   344 &  5,423 \\
	$1M$ & $100M$  &   880 & 3,224 & 43,806 \\
	$10M$ & $100M$ & 1,366 & 3,360 & 54,171 \\
        \hline
\end{tabular}
}
\label{table:exp_memory}
\end{table}

Table~\ref{table:exp_memory} shows the results. Notice, that SNAP can store a graph of 10M nodes, and 100M edges in mere 1.3GB of memory, while iGraph needs over 3.3GB and NetworkX requires nearly 55GB of memory to store the same graph.
It is somewhat surprising that iGraph
	requires about 3 times more memory than SNAP,
	despite using vectors to represent nodes
		rather than a hash table.
NetworkX uses hash tables extensively and
	it is thus not surprising that
		it requires over 40 times more memory than SNAP.

We used the memory consumption measurements in Table~\ref{table:exp_memory} to calculate the number of bytes required by each library to represent a node or an edge.
As can be seen in Table~\ref{table:exp_bytes}, SNAP requires four times less memory per edge than iGraph and 50 times less memory per edge than NetworkX. Since graphs have usually significantly more edges than nodes, memory requirements to store the edges are the main indicator of the size of graphs that will fit in a given amount of RAM. 

\begin{table}[h]
\centering
\tbl{Memory requirements to represent a node or an edge, based on the measurements of the $G(n,m)$ model. Memory usages are in bytes. SNAP uses four times less memory per edge than iGraph and over 50 times less memory per edge than NetworkX.}
{
\begin{tabular}{r | r r r}
        \hline
        \multicolumn{1}{c|}{} & \multicolumn{3}{c}{Memory usage [bytes]}  \\
        Item & SNAP & iGraph & NetworkX  \\
        \hline
	Node & 54.4 & 24.0 & 1158.2 \\
	Edge &  8.3 & 32.0 &  426.5 \\
        \hline
\end{tabular}
}
\label{table:exp_bytes}
\end{table}

\begin{figure}[t]
\centering
\includegraphics[width=0.75\columnwidth]{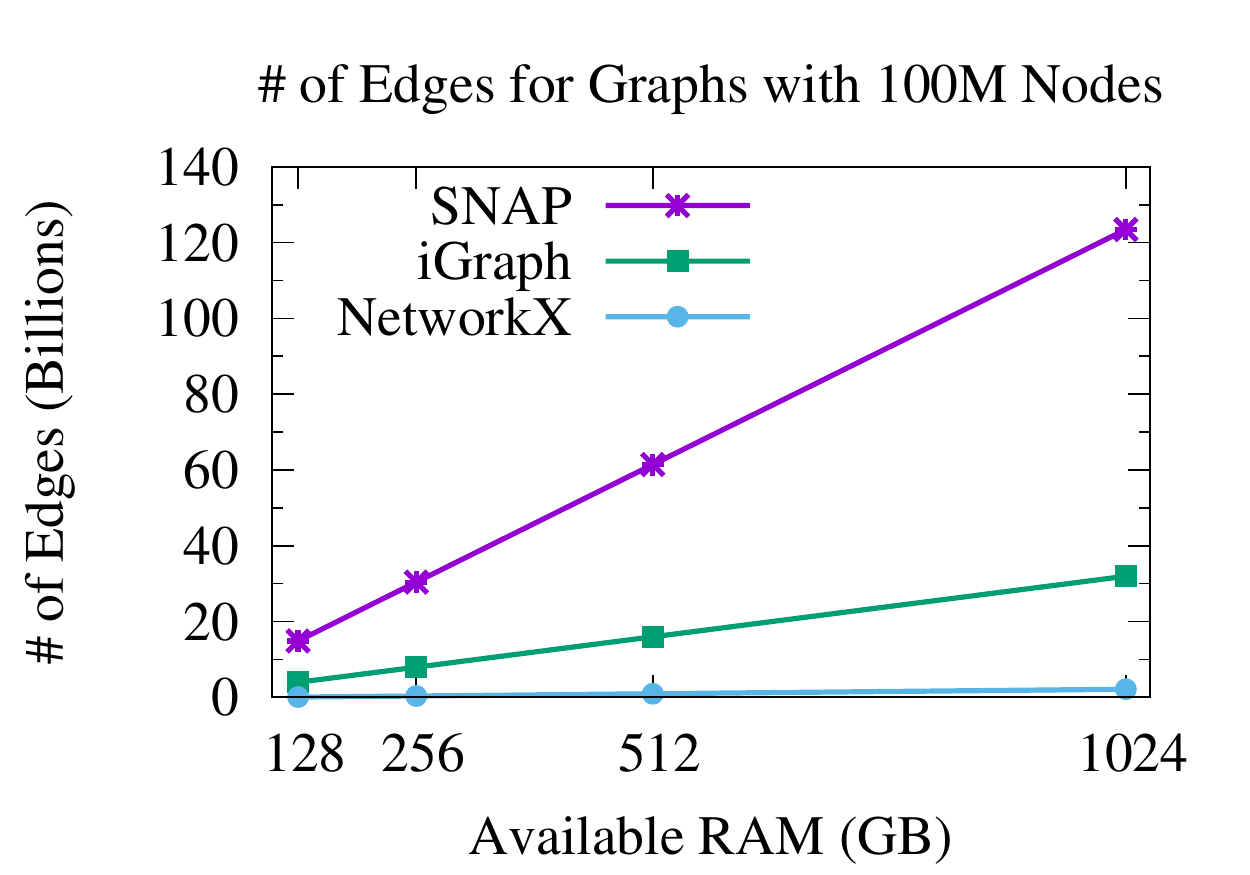}
\caption{Maximum graph sizes for varying RAM availability. Number of nodes is fixed at 100 million, estimated maximum number of edges is shown. Using 1TB RAM, SNAP can fit over 120 billion edges, iGraph 30 billion, and NetworkX 2 billion.}
\label{figure:ram}
\end{figure}

We illustrate the size of a graph that can be represented by each system in a given amount of RAM by fixing the number of nodes at 100 million and then calculating the maximum number of edges that fit in the remaining RAM, using numbers from Table~\ref{table:exp_bytes}. The results are shown in Figure~\ref{figure:ram}. For 1024GB of RAM, SNAP can represent graphs with 123.5 billion edges, iGraph 31.9 billion edges, and NetworkX 2.1 billion edges.

\subsection{Basic Graph Operations}

Next, we measure execution times of basic graph operations
	for an Erd\H{o}s-R\'{e}nyi random graph $G(1M,100M)$.

\begin{table}[t]
\centering
\tbl{Execution times for basic graph operations on Erd\H{o}s-R\'{e}nyi random graph $G(1M,100M)$. Times are in seconds. Overall, SNAP is about two times slower than iGraph at generating the graph but it is 15 times faster at loading and saving it to the disk. NetworkX is 5 to 200 times slower than SNAP.}
{
\begin{tabular}{r | r r r}
        \hline
        & \multicolumn{3}{c}{Execution time [seconds]} \\
        Operation & SNAP & iGraph & NetworkX  \\
        \hline
	Generate   & 139.3 & 74.2 & 748.7 \\
	Save       &   3.3 & 47.0 & 757.2 \\
	Load       &   4.6 & 87.8 & 522.0 \\
        \hline
\end{tabular}
}
\label{table:exp_basic}
\end{table}

First, we examine the times for generating a graph,
    saving the graph to a file, and
    loading the graph from the file.
Results are shown in Table~\ref{table:exp_basic}.
We used a built-in function in each system to generate the graphs.
For graph generation, SNAP is about two times slower than iGraph, and
	more than 5 times faster than NetworkX (Table~\ref{table:exp_basic}). However, graph generation
    in SNAP is inserting one edge at a time, while iGraph has an optimized
    implementation that inserts edges in bulk.

The performance of graph loading and saving operations is often a bottleneck in graph analysis.
For these operations,
	SNAP is over 15 times faster than iGraph and
	100 times faster than NetworkX (Table~\ref{table:exp_basic}).
The benchmark utilized an internal binary representation of graphs for SNAP,
	while a text representation was used for iGraph and NetworkX.
SNAP and iGraph have similar performance when saving/loading graphs from/to a textual format. So, the advantage of SNAP over iGraph can be attributed
		to the SNAP support for the binary graph representation on the disk.

\begin{table}[t]
\centering
\tbl{Testing edge existence. Edges are random, the number of tests is equal to the number of total edges in the graph. Times are in seconds. SNAP is about 10-20\% faster than or comparable to iGraph, while being 3-5 times faster than NetworkX.}
{
\begin{tabular}{r r | r r r}
        \hline
        \multicolumn{2}{c|}{Graph size} & \multicolumn{3}{c}{Execution time [seconds]} \\
        Nodes & Edges & SNAP & iGraph & NetworkX  \\
        \hline 
        $1M$ & $10M$ &  3.8 &  5.2  &  23.5 \\
        $1M$ & $100M$ & 75.4 & 113.8 & 218.3 \\
        $10M$ & $100M$ & 67.9 &  63.3 & 255.8 \\
        \hline
\end{tabular}
}
\label{table:exp_edge}
\end{table}

Second, we also benchmark the fundamental operations when working with graphs. We focus on the time it takes to test for the existence of a given edge $(i,j)$. We performed an experiment where we generated larger and larger instances of Erd\H{o}s-R\'{e}nyi random graphs and measured execution times for testing the presence of edges in a given graph. For each test, we generated a random source and destination node and tested for its existence in the graph. The number of test iterations is equal to the number of edges in the graph. Table~\ref{table:exp_edge} gives the results and we notice that SNAP is about 10-20\% faster than or comparable to iGraph and 3-5 times faster than NetworkX.

\begin{table}[t]
\centering
\tbl{Execution times for deleting 10\% of nodes and their corresponding edges from Erd\H{o}s-R\'{e}nyi random graph $G(1M,10M)$. Times are in seconds. SNAP is four to five times faster than iGraph and NetworkX. However, if one deletes nodes from the graph one-by-one in iGraph, its performance slows down for five orders of magnitude.}
{
\begin{tabular}{r | r r r}
        \hline
        & \multicolumn{3}{c}{Execution time [seconds]} \\
        Operation & SNAP & iGraph & NetworkX  \\
        \hline 
	Deleting nodes     & 0.7 & 3.0 & 4.1 \\
        \hline
\end{tabular}
}
\label{table:exp_del}
\end{table}

Last, we also estimate system flexibility,
    which tells us how computationally expensive it is
        to modify graph structure,
    by measuring the execution times of deleting 10\% of nodes and
        their corresponding edges from $G(1M,10M)$.
SNAP is much faster than iGraph and NetworkX
	when deleting nodes from the graph (Table~\ref{table:exp_del}).
Furthermore,
	the nodes in SNAP and NetworkX were deleted incrementally,
	one node at the time,
	while the nodes in iGraph were deleted
		in a single batch with one function call.
When nodes were deleted one by one in iGraph as well,
	it took 334,720 seconds to delete 10\% of nodes in the graph.
The fact that SNAP is more than 5 orders of magnitude faster than iGraph indicates that iGraph's graph data structures are optimized for speed on static graphs while also being less memory efficient. However, the iGraph data structure seems to completely fail in case of dynamic graphs where nodes/edges appear/disappear over time.

\subsection{Graph Algorithms}

To evaluate system performance on a real-world graph,
		we used a friendship graph
		of the LiveJournal online social network~\cite{snapnets}.
The LiveJournal network has about 4.8M nodes and 69M edges.
We measured execution times for common graph analytics operations:
	PageRank, clustering coefficient, weakly connected components,
	extracting 3-core of a network, and testing edge existence.
For the PageRank algorithm,
	we show the time it takes to perform 10 iterations of the algorithm.

\begin{table}[t]
\centering
\tbl{Execution times for graph algorithms on the LiveJournal network with 4.8M nodes and 69M edges. Times are in seconds. Generally we observe that due to a hash based graph representation that allows efficient changes in the structure, SNAP is equal to iGraph in some graph operations while about 3 times slower in algorithms that benefit from fast vector access in iGraph. NetworkX is much slower than either SNAP or iGraph in most operations.}
{
\begin{tabular}{r | r r r}
        \hline
        & \multicolumn{3}{c}{Execution time [seconds]} \\
        Operation & SNAP & iGraph & NetworkX  \\
        \hline 
	PageRank               &  40.9 & 10.6 & 2,720.8 \\
	Clustering Coefficient & 143.3 & 58.5 & 4,265.4 \\
	Connected Components   &  13.3 &  5.8 &    60.3 \\
	3-core                 &  37.9 & 41.7 & 2,276.1 \\
	Test Edge Existence    &  45.7 & 35.2 &   158.6 \\
        \hline
\end{tabular}
}
\label{table:exp_algs}
\end{table}

Table~\ref{table:exp_algs} gives the results.
We can observe that SNAP is only about 3 times slower than iGraph in some operations and about equal in others,
	while it is between 4 to 60 times faster than NetworkX (Table~\ref{table:exp_algs}).
As expected, NetworkX performs the best when the algorithms require
	mostly a large number of random accesses
		for which hash tables work well,
	while it performs poorly when the algorithm execution
		is dominated by sequential data accesses
			where vectors dominate.

\vspace{2mm}
In summary, we find that the SNAP graph data structure is by far the most memory efficient and also most flexible as it is able to add/delete nodes and edges the fastest. In terms of input/output operations SNAP also performs the best. And last, we find that SNAP offers competitive performance in executing static graph algorithms.

\subsection{Comparison to Distributed Graph Processing Frameworks}

So far we focused our experiments on SNAP performance on a sequential execution of a single thread on a single machine. However, we have also been studying how to extend SNAP to single machine multi-threaded architectures.

We have implemented parallel versions of several SNAP algorithms. Our experiments have shown that a parallel SNAP on a single machine can offer comparable performance to specialized algorithms and even frameworks utilizing distributed systems for network analysis and mining~\cite{perez15ringo}. Results are summarized in Table~\ref{tab:exp_parallel}. For example, triangle counting on the Twitter2010 graph~\cite{Twitter2010}, which has about 42 million nodes and 1.5 billion edges, required 469s on a 6 core machine~\cite{kim2014opt}, 564s on a 200 processor cluster~\cite{Patric}, while the parallel SNAP engine on a single machine with 40 cores required 263s.

\begin{table}[h]
\centering
\tbl{Execution times for graph algorithms on the Twitter2010 network with 42M nodes and 1.5B edges. Times are in seconds.}
{
\begin{tabular}{l|lc}
\hline
Benchmark & System & Execution time [seconds] \\
\hline
\multirow{3}{*}{Triangles} & \multicolumn{1}{l}{OPT, 1 machine, 6 cores~\cite{kim2014opt}} & \multicolumn{1}{c}{469} \\
	 & \multicolumn{1}{l}{PATRIC, 200 processor cluster~\cite{Patric}} & \multicolumn{1}{c}{564} \\
	 & \multicolumn{1}{l}{SNAP, 1 machine, 40 cores} & \multicolumn{1}{c}{263} \\
\hline
\multirow{2}{*}{PageRank} & \multicolumn{1}{l}{PowerGraph, 64 machines, 512 cores~\cite{gonzalez2012powergraph}} & \multicolumn{1}{c}{3.6} \\
	 & \multicolumn{1}{l}{SNAP, 1 machine, 40 cores} & \multicolumn{1}{c}{6.0} \\
\hline
\end{tabular}
}
\label{tab:exp_parallel}
\end{table}

We obtained similar results by measuring execution time of the PageRank algorithm~\cite{PageRank} on the same graph. PowerGraph~\cite{gonzalez2012powergraph}, a state-of-the-art distributed system for network analysis running on 64 machines with 512 cores, took 3.6s per PageRank iteration, while our system needed 6s for the same operation using only one machine and 40 cores, a significantly simpler configuration and more than 12 times fewer cores.

Note also that SNAP uses only about 13GB of RAM to process the Twitter2010 graph, so the graph fits easily in the RAM of most modern laptops.

These results, together with the sizes of networks being analyzed, demonstrate that a single multi-core big-memory machine provides an attractive platform for network analysis of a large majority of networks~\cite{perez15ringo}.


\section{Stanford Large Network Dataset Collection}
\label{sec:data}

As part of SNAP, we are also maintaining and making publicly available the Stanford Large Network Dataset Collection~\cite{snapnets}, a set of around 80 different social and information real-world networks and datasets from a wide range of domains, including social networks, citation and collaboration networks, Internet and Web based networks, and media networks.
Table~\ref{tab:data} gives the types of datasets in the collection.

\begin{table}
\small
\centering
\tbl{Datasets in the Stanford Large Network Dataset Collection.}
{
\begin{tabular}{lcl}
\hline
	{\bf Dataset type} & {\bf Count} & {\bf Sample datasets} \\
	\hline
	{\it Social networks} & 10 &
		Facebook, Google+, Slashdot, Twitter, Epinions \\
	{\it Ground-truth communities} & 6 & 
		LiveJournal, Friendster, Amazon products \\
	{\it Communication networks} & 3 & 
		Email, Wikipedia talk \\
	{\it Citation networks} & 3 & 
		Arxiv, US patents \\
	{\it Collaboration networks} & 5 & 
		Arxiv \\
	{\it Web graphs} & 4 & 
		Berkeley, Stanford, Notre Dame \\
	{\it Product co-purchasing networks} & 5 & 
		Amazon product \\
	{\it Internet peer-to-peer networks} & 9 & 
		Gnutella \\
	{\it Road networks} & 3 & 
		California, Pennsylvania, Texas \\
	{\it Autonomous systems graphs} & 5 & 
		AS peering, CAIDA, Internet topology \\
	{\it Signed networks} & 6 & 
		 Epinions, Wikipedia, Slashdot Zoo \\
	{\it Location-based social networks} & 2 & 
		 Gowalla, Brightkite \\
	{\it Wikipedia networks} & 6 & 
		 Navigation, voting, talk, elections, edit history \\
	{\it Memetracker and Twitter} & 4 & 
		 Post hyperlinks, popular phrases, tweets \\
	{\it Online communities} & 2 & 
		 Reddit, Flickr \\
	{\it Online reviews} & 6 & 
		 BeerAdvocate, RateBeer, Amazon, Fine Foods \\
	\hline
\end{tabular}
}
	\label{tab:data}
\end{table}

The datasets were collected as part of our research in the past and in that sense represent typical graphs being analyzed.
Table~\ref{tab:datasize} gives the distribution of graph sizes in the collection.
It can be observed that a vast majority of graphs are relatively small with less than 100 million edges and thus can easily be 
analyzed in SNAP.
The performance benchmarks in Table~\ref{table:exp_algs} are thus indicative of the execution times of graph algorithms being applied to real-world networks.

\begin{table}
\centering
\tbl{Distribution of graph sizes in the Stanford Large Network Dataset Collection.}
{
\begin{tabular}{c r}
\hline
Graph size & Number of \\
(number of edges) & graphs \\
\hline
\textless 0.1M &  18 \\
0.1M -- 1M &  24 \\
1M -- 10M &  17 \\
10M -- 100M &   7 \\
100M -- 1B &   4 \\
\textgreater 1B &   1 \\
\hline
\end{tabular}
}
\label{tab:datasize}
\end{table}


\section{Resources}
\label{sec:documentation}

SNAP resources are available from our Web site at: \texttt{http://snap.stanford.edu}.

The site contains extensive user documentation, tutorials, regular SNAP stable releases, links to the relevant GitHub repositories, a programming guide, and the datasets from the Stanford Large Network Dataset Collection.

Complete SNAP source code has been released under
	a permissive BSD type open source license.
SNAP is being actively developed.
We welcome community contributions to the SNAP code base and the SNAP dataset collection.


\section{Conclusion}
\label{sec:conclusion}

We have presented SNAP, a system for analysis of large graphs.
We demonstrate that graph representation employed by SNAP
	is unique in the sense that 
	it provides an attractive balance between
	the ability to efficiently modify graph structure and
	the need for fast execution of graph algorithms.
While SNAP implements efficient operations
	to add or delete nodes and edges in a graph,
	it imposes only limited overhead on graph algorithms.
An additional benefit of SNAP graph representation is that it is compact and
	requires lower amount of RAM than alternative representations,
	which is useful in analysis of large graphs.

We are currently extending SNAP in several directions.
One direction is speeding up algorithms via parallel execution.
Modern CPUs provide a large number of cores,
	which provide a natural platform for parallel algorithms.
Another direction is exploring ways of
	how the graphs are constructed from data and
	then identify powerful primitives
	that cover a broad range of graph construction scenarios.

\acks{
Many developers contributed to SNAP. Top 5 contributors to the repository, excluding the authors, are Nicholas Shelly, Sheila Ramaswamy, Jaewon Yang, Jason Jong, and Nikhil Khadke.
We also thank Jo\v{z}ef Stefan Institute for making available their GLib library.
}


\vskip 0.2in
\bibliographystyle{ACM-Reference-Format-Journals}
\bibliography{refs}

\end{document}